\documentclass{IEEEtran}
\usepackage{color}
\usepackage{}
\usepackage{amsmath}
\usepackage{amsfonts}
\usepackage{graphicx} 
\usepackage{caption}
\usepackage{subcaption} 
\usepackage{enumerate}
\usepackage[round]{natbib}
\usepackage{hyperref}
\usepackage[capitalise]{cleveref}
\usepackage{url}
\usepackage{authblk}

\usepackage{xcolor}
\hypersetup{
    colorlinks = true,
    linkcolor = cyan,
    linkbordercolor = {white},
    citecolor=blue,
}

\newcommand{\ie}[0]{\emph{i.e.},~}
\newcommand{\eg}[0]{\emph{e.g.},~}

\newcommand{\etc}{\emph{etc.~}}

\newcommand{\defeq}{\ensuremath{\overset{\text{def}}{=}}}
\newcommand{\yv}[0]{\ensuremath{{y}}}
\newcommand{\xv}[0]{\ensuremath{{x}}}
\newcommand{\zv}[0]{\ensuremath{{z}}}

\newcommand{\xvh}[0]{\ensuremath{\hat{x}}}
\newcommand{\yvh}[0]{\ensuremath{\hat{y}}}
\newcommand{\ZV}[0]{\ensuremath{{z}}}
\newcommand{\YY}[0]{\ensuremath{\mathcal{Y}}}
\newcommand{\XX}[0]{\ensuremath{\mathcal{X}}}
\newcommand{\ZZ}[0]{\ensuremath{\mathcal{Z}}}
\newcommand{\DD}[0]{\ensuremath{\mathcal{D}}}
\newcommand{\Dist}[0]{\ensuremath{\mathbb{D}_{\mathsf{KL}}}}

\newcommand{\phis}[0]{\ensuremath{\boldsymbol{\phi}}}
\newcommand{\Ex}[0]{\ensuremath{\mathbb{E}}}
\newcommand{\ttt}[1]{\ensuremath{^{(#1)}}}

\newcommand{\COSMOS}[0]{{\textsc{cosmos}~}}
\newcommand{\GALAXYZOO}[0]{{\textsc{galaxy-zoo}~}}
 \begin{document}
%
\title{Enabling Dark Energy Science with Deep Generative Models of Galaxy Images}
\renewcommand\Affilfont{\itshape\small}
\author[1]{Siamak Ravanbakhsh}
\author[2]{Fran\c{c}ois Lanusse}
\author[2]{Rachel Mandelbaum}
\author[1]{Jeff Schneider}
\author[1]{Barnab\'as P\'oczos}
\affil[1]{School of Computer Science, Carnegie Mellon University}
\affil[2]{McWilliams Center for Cosmology, Carnegie Mellon University}

\maketitle
\begin{abstract}
Understanding the nature of dark energy, the mysterious force driving the accelerated expansion of the Universe, is a major challenge of modern cosmology. The next generation of cosmological surveys, specifically designed to address this issue, rely on accurate measurements of the apparent shapes of distant galaxies. However, shape measurement methods suffer from various unavoidable biases and therefore will rely on a precise calibration to meet the accuracy requirements of the science analysis. This calibration process remains an open challenge as it requires large sets of high quality galaxy images. 
To this end, we study the application of deep conditional generative models in generating realistic galaxy images.
In particular we consider variations on conditional variational autoencoder and introduce a new adversarial objective for training of conditional generative networks.
Our results suggest a reliable alternative to the acquisition of expensive high quality observations for generating the calibration data needed by the next generation of cosmological surveys.
\end{abstract}


\noindent The last two decades have greatly clarified the contents of the Universe, while leaving several large mysteries in our cosmological model. We now have compelling evidence that the expansion rate of the Universe is accelerating, suggesting that the vast majority of the total energy content of the Universe is the so-called \textit{dark energy}. Yet we lack an understanding of what dark energy actually is, which provides one of the main motivations behind the next generation of cosmological surveys such as LSST \cite{LSST2009}, Euclid \cite{Laureijs2011} and WFIRST \cite{Green2012}. These billion dollar projects are specifically designed to shed light  on the nature of dark energy  by probing the Universe through the \textit{weak gravitational lensing} effect --\ie the minute deflection of the light from distant objects by the intervening massive large scale structures of the Universe. On cosmological scales, this lensing effect causes very small but coherent deformations of background galaxy images, which appear slightly \textit{sheared}, providing a way to statistically map the matter distribution in the Universe. To measure the lensing signal, future surveys will image and measure the shapes of billions of galaxies, significantly driving down statistical errors compared to the current generation of surveys, to the level where dark energy models may become distinguishable.

However, the quality of this analysis hinges on the accuracy of the shape measurement algorithms tasked with estimating the \textit{ellipticities} of the  galaxies in the survey. This point is particularly crucial to the success of  these missions, as any unaccounted for measurement biases in their ensemble averages would impact the final cosmological analysis and potentially lead to false conclusions. In order to detect and/or calibrate any such biases, future surveys will heavily rely on image simulations, closely mimicking real observations but with a known ground truth lensing signal.

\begin{figure}[h]
\centering
\includegraphics[width=1\linewidth]{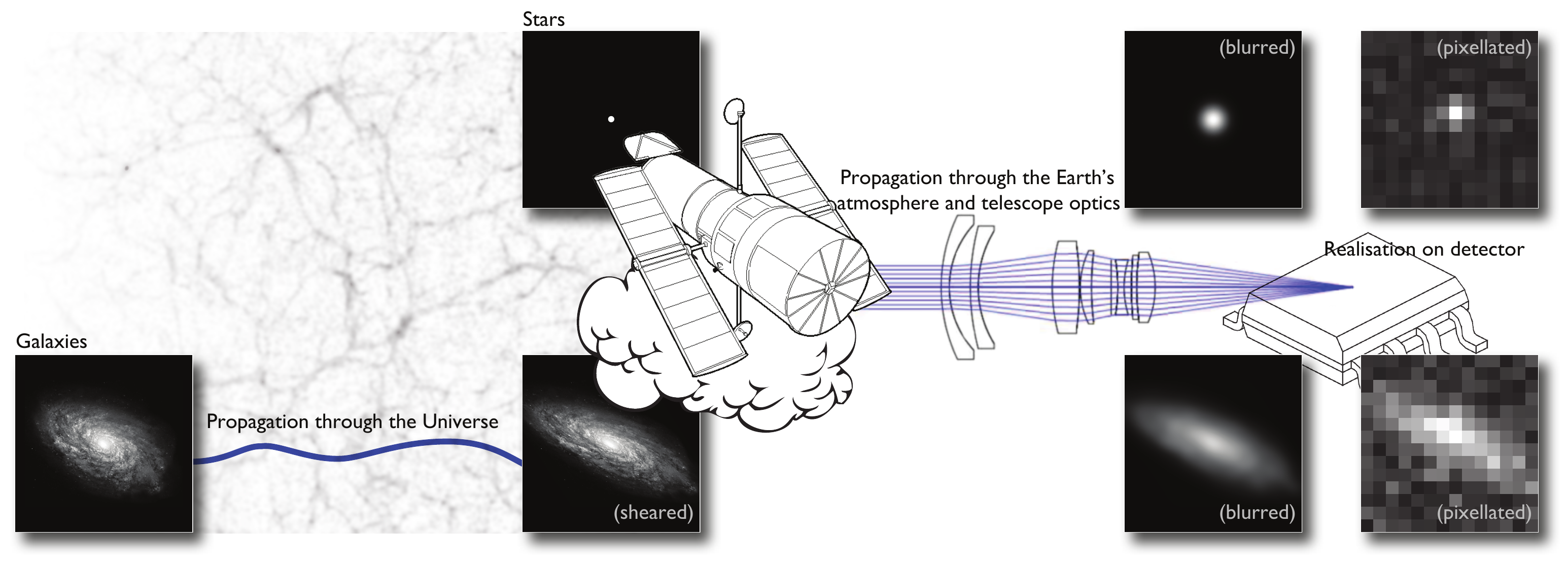}
\caption{{\small Illustration of the processes involved in the measurement of weak gravitational lensing. The light from distant galaxies is deflected by the matter in the Universe, causing a shearing of the galaxy images, which are then further blurred by the atmosphere and the telescope optics and finally pixelated into a noisy image by the imaging sensor. Image credit: \citet{Mandelbaum2014}, adapted from \citet{Kitching2010}.}}
\label{fig:great}
\end{figure}

Producing these image simulations, however, is challenging in itself as they require high quality galaxy images as the input of the simulation pipeline. Such observations can only be obtained by extremely expensive space-based imaging surveys, which will remain a scarce resource for the foreseeable future. The largest current survey being used for image simulation purposes is the \COSMOS survey \cite{Scoville2007}, carried out using the \textit{Hubble Space Telescope} (HST). Despite being the largest available dataset, \COSMOS is relatively small, and there is great interest in increasing the size of our galaxy image samples to improve the quality of this crucial calibration process.

\begin{figure*}
  \centering
\includegraphics[width=1\linewidth]{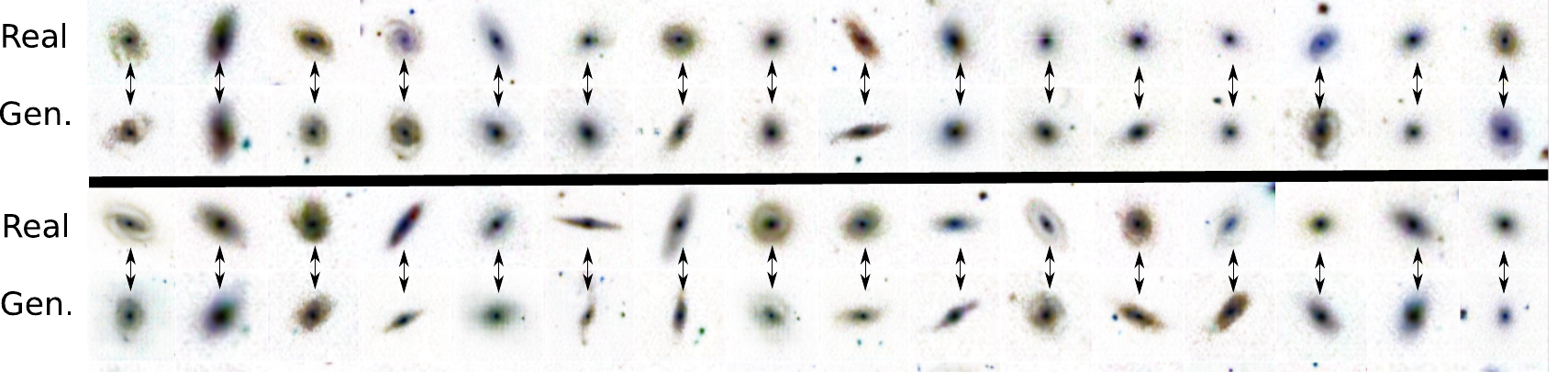} 
  \caption{{\small Samples from the \GALAXYZOO dataset versus generated samples using conditional generative adversarial network of \cref{sec:cgan}. Each synthetic image is a $128 \times 128$ colored image (here inverted) produced by conditioning on a set of features $y \in [0,1]^{37}$. The pair of observed and generated images in each column correspond to the same $y$ value. For details on these crowd-sourced $y$ features see \cite{willett2013galaxy}. 
  These instances are selected from the test-set and were unavailable to the model during the training.}}
\label{fig:galaxyzoo}
\end{figure*}

In this work, we propose an alternative to the expensive acquisition of more high quality calibration data using deep conditional generative models. 
In recent years, these models have achieved remarkable success in modeling complex high-dimensional distributions, producing natural images that can pass the visual Turing test.
Two prominent approaches for training these models are \textbf{variational autoencoder} (VAE) \cite{kingma2013auto,rezende2014stochastic} and \textbf{generative adversarial network} (GAN)~\cite{goodfellow2014generative}.
Our aim is to train a coditional variation of these models using existing HST data 
and generate new galaxy images ``conditioned'' on statistics of interest such as the brightness or size of the galaxy. This will allow us to synthesize calibration datasets for specific galaxy populations, with objects exhibiting realistic morphologies. 
In related works in machine learning literature \citet{regier2015celeste} use a convex combination of smooth and spiral templates in an (unconditioned) generative model of galaxy images and \citet{deep_galaxy} propose using VAE for this task.\footnote{The current approach to address this problem in cosmology literature is to fit analytic parametric light profiles (defined by size, intensity, ellipticity and steepness parameters) to the observed galaxies, followed by a simple modelling of the distribution of the fitted parameters as a function of a quantity of interest, such as the galaxy brightness. This modelling usually simply involves fitting a linear dependence of mean and standard deviation of a Gaussian distribution -- \eg see \cite{hoekstra2016study}; Appendix A. However, simple parametric models of galaxy light profiles do not have the complex morphologies needed for calibration task. The only currently available alternative, if realistic galaxy morphologies are needed, is to use the training set images themselves as the input of the simulation pipeline. This involves subsampling the training set to match the distribution of size, redshift and brightness of the target galaxy simulations, leaving only a relatively small number of objects, reused several hundred times to simulate a large survey -- \eg see \cite{jarvis2016science}; Section 6.1. }





In the following, \cref{sec:calibration} gives a brief background on the image generation for calibration and its significance for modern cosmology.
We then review the current approaches to deep conditional generative models and introduce new techniques for our problem setting in \cref{sec:cvae,sec:cgan}.
In \cref{sec:validation} we assess the quality of the generated images by comparing the conditional distributions of shape and morphology parameters between simulated and real galaxies, and find good agreement.

\section{Weak Gravitational Lensing}\label{sec:calibration}
In the weak regime of gravitational lensing, the distortion of background galaxy images can be modeled by an anisotropic shear, noted $\gamma$, whose amplitude and orientation depend on the matter distribution between the observer and these distant galaxies. This shear affects in particular the apparent ellipticity of galaxies, denoted $e$. 
Measuring this weak lensing effect is made possible under the assumption that background galaxies are randomly oriented, so that the ensemble average of the shapes would average to zero in the absence of lensing.  Their apparent ellipticity $e$ can then be used as a noisy but unbiased estimator of the shear field $\gamma$: $\Ex [ e ] = \gamma$. The cosmological analysis then involves computing auto- and cross-correlations of the measured ellipticities for galaxies at different distances. These correlation functions are compared to theoretical predictions in order to constrain cosmological models and shed light on the nature of dark energy.

However, measuring galaxy ellipticities such that their ensemble average (used for the cosmological analysis) is unbiased  is an extremely challenging task. \cref{fig:great} illustrates the main steps involved in the acquisition of the science images. The weakly sheared galaxy images undergo additional distortions (essentially blurring) as they go through the atmosphere and telescope optics, before being acquired by the imaging sensor which pixelates the noisy image. As this figure illustrates, the cosmological shear is clearly a subdominant effect in the final image and needs to be disentangled from subsequent blurring by the atmosphere and telescope options.  This blurring, or Point Spread Function (PSF),  can be directly measured by using stars as  point sources, as shown at  the top of~\cref{fig:great}. 

\begin{figure*}
  \centering
\includegraphics[width=1\linewidth]{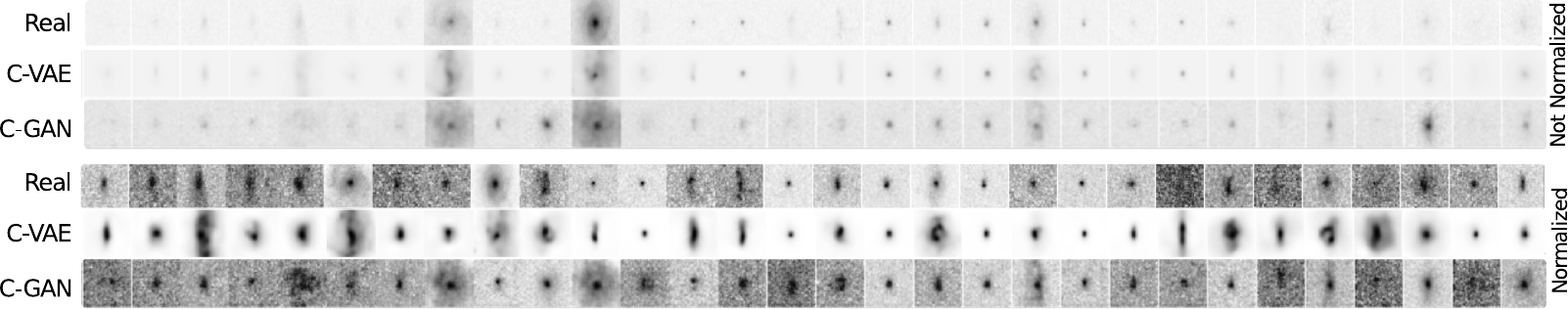} 
  \caption{{\small Samples from the \COSMOS dataset and generated samples using the 
  conditional variational autoencoder (C-VAE, scheme I) and
  our variation on conditional generative adversarial network (C-GAN). Each column image shows three $64 \times 64$ images (here inverted) produced by conditioning on the same set of features $y \in \Re^{3}$ in the test-set. Due to its high dynamic range, most figures are very faint. In the bottom three rows, each image is individually normalized.}}
\label{fig:cosmos_images}
\end{figure*}

Once the image is acquired, shape measurement algorithms are used to estimate the ellipticity of the galaxy while correcting for the PSF. However, despite the best efforts of the weak lensing community for nearly two decades, all current state-of-the-art shape measurement algorithms are still susceptible to biases in the inferred shears. These  measurement biases are commonly modeled in terms of additive and multiplicative bias parameters $c$ and $m$ defined as:
\begin{equation}
	\Ex[e] = (1 + m) \ \gamma  + c
	\label{eq:shear_bias}
\end{equation}
where $\gamma$ is the true shear. Depending on the shape measurement method being used, $m$ and $c$ can depend on factors such as the PSF size/shape, the level of noise in the images or, more generally, intrinsic properties of the galaxy population (like their size and ellipticity distributions, \etc).
Calibration of these biases can be achieved using image simulations, closely mimicking real observations for a given survey but using galaxy images distorted with a known shear, thus allowing the measurement of the bias parameters in \cref{eq:shear_bias}.

\textbf{Image simulation pipelines}, such as the \textit{GalSim} package \cite{Rowe2015}, use a forward modeling of the observations, reproducing all the steps of the image acquisition process in \cref{fig:great}, and therefore require as a starting point galaxy images with high resolution and S/N. The main difficulty in these image simulations is therefore the need for a calibration sample of high quality galaxy images representative of the galaxy population of the survey being simulated. Our aim in this work is to train a deep generative model which can be used to cheaply synthesize such data sets for specific galaxy populations, by conditioning the samples on measurable quantities. 

\subsection{Data set}\label{sec:dataset}
As our main dataset, we use the \COSMOS survey to build a training and validation set of galaxy images and extract from the corresponding catalog a condition vector $y$ with three features: half-light radius (measure of size), magnitude (measure of brightness) and redshift (cosmological measure of distance). To facilitate the training, we align all galaxies along their major axis and produce 85,000 instances of 64x64 image stamps using the GalSim package. 

We also use the \GALAXYZOO dataset~\cite{willett2013galaxy} to demonstrate the abilities of our alternative conditional adversarial objective. Each of the 61,000 galaxy images in this dataset is accompanied by $\yv \in [0,1]^{37}$ features produced using a crowd-sourced set of questions that form a decision tree. We cropped the central $50\%$ of these images and resized them to $128\times128$ pixels. We augmented both datasets by flipping the images along the vertical and horizontal axes.

\section{Conditional Variational Autoencoder}\label{sec:cvae}
Applications in semi-supervised learning and structured prediction have motivated different versions of the ``conditional'' variational autoencoder (C-VAE) in the past \cite{kingma2014semi,sohn2015learning}. Although the  architecture that we discuss here resembles to those of \cite{kingma2014semi,sohn2015learning}, there are some differences due to different objectives.

We are interested in learning the conditional density $p^*(\xv \mid \yv)$ for $\xv \in \XX$ and $\yv \in \YY$, given a set of observations $\DD = (\xvh^1,\yvh^1),\ldots,(\xvh^N, \yvh^N)$, by learning model parameters $\theta$ that maximizes the conditional likelihood $\prod_{(\xvh, \yvh) \in \DD} p_\theta(\xvh \mid \yvh)$ -- \eg for the \COSMOS dataset $\XX = \Re^{64 \times 64}$ and $\YY = \Re^3$.
In a \textit{latent-variable model}, an auxiliary variable $\zv \in \ZZ$ is introduced to increase the expressive power of $p_\theta(\xv, \zv \mid \yv)$, such that $\int_{\ZZ} p_\theta(\xv, \zv \mid \yv) \mathrm{d}z$ is the marginal of interest. Here, different assignments to $\zv$ can explain variations and complex statistical dependencies in $p(\xv \mid \yv)$.

To enable efficient (ancestral) sampling from this model, $p_\theta$ can be a directed model $p_\theta(\xv, \zv \mid \yv) = p_{\theta_1}(\zv \mid \yv)\, p_{\theta_2}(\xv \mid \zv, \yv)$, where we first sample 
$\zv \sim p_{\theta_1}(\cdot \mid \yv)$ followed by $\xv \sim p_{\theta_2}(\cdot \mid \zv, \yv)$.
An expressive form for the conditional distributions $p_{\theta_1}$ and $p_{\theta_2}$ is a deep neural network, that can represent complex directed graphical models. Here, for example, we model $p_{\theta_2}(\xv \mid \zv, \yv)$ using multi-layered convolutional or densely connected neural networks that encode the mean and variance of a multi-variate Gaussian for the \COSMOS dataset and the expectation of Bernoulli variables for the \GALAXYZOO dataset. 
 
To learn the parameters $\theta$ one needs to estimate the posterior $p_{\theta}(\zv \mid \xv, \yv)$, which is often intractable in directed models. An elegant solution to this problem is to introduce a second directed model $q(\zv \mid \xv, \yv)$, called inference or \textit{recognition model}. This conditional distribution is also encoded as a deep neural network and it is tasked with estimating the intractable posterior $p_{\theta}(\zv \mid \xv, \yv)$.

This is achieved through a \textbf{variational bound} on the conditional log-likelihood:
\begin{align}
  \label{eq:nll_uncond}
  \log(p_{\theta}(\xvh \mid \yvh)) \geq  
 &-\Dist(q_{\phis}(\zv \mid \xvh, \yvh) \|  p_{\theta_1}(\zv \mid \yvh))\\ &+ \Ex_{\ZV \sim q_{\phis}(\cdot \mid \xvh, \yvh)}[\log p_{\theta_2}(\xvh \mid \zv, \yvh) ] \notag
\end{align}
where the first term is the KL-divergence between the posterior $q_{\phis}$ and the conditional prior  $p_{\theta_1}$ and the second term is the reconstruction error -- that is we want the model to achieve low reconstruction error
while encoding the dataset. At the same time the KL-divergence term encourages the code to follow a distribution, dictated by the the condition $\yv$. 
Fortunately, the reparametrization-trick by \cite{kingma2013auto,rezende2014stochastic,williams1992simple} enables the maximization of this lower-bound (\ie learn $\theta_1$, $\theta_2$ and $\phis$) using stochastic back-propagation through the layers of these three neural networks. This enables maximizing  the log-likelihood
of an expressive model with large number of parameters through variations of stochastic gradient descent. 

\begin{figure}
  \centering
\includegraphics[width=1\linewidth]{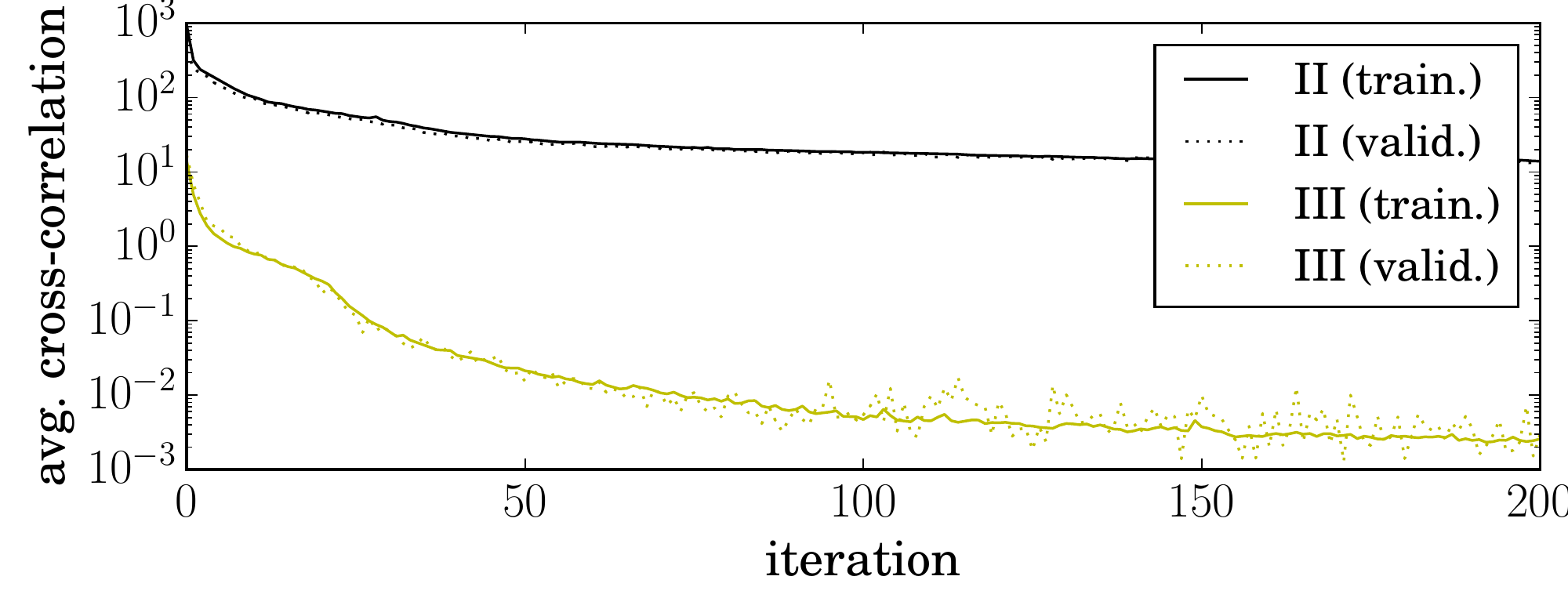} 
  \caption{{\small Cross-correlation between $y$ and $z$ in C-VAE when $p(z \mid y) = p(z)$, with and without cross-correlation penalty.}}\label{fig:cc}
\end{figure}

\subsection{Cross-Correlation}
Inspired by the application of cross-correlation in disentangling the factors in an autoencoder by \citet{cheung2014discovering}, we also consider an alternative method of conditioning in VAE.
Let us proceed with a simple question:
what happens here if we simplify the prior $p_{\theta_1}(\zv \mid \yv) \Rightarrow p_{\theta_1}(\zv)$?  
In principle, the simplified C-VAE would try to make the posterior $q_\phi(\zv \mid \xv, \yv)$
 independent of $\yv$.\footnote{This is because the information content of $\yv$ is already available to the generative model $p_\theta(x \mid y,z)$ for reconstruction
and reducing the information exchange through $z$ should reduce the KL-divergence penalty $\Dist(q_{\phis}(\zv \mid \xvh, \yvh) \|  p(\zv))$.}
In this case, for generating samples $\xv \sim p_\theta(\cdot \mid \yv)$, we could still sample $\zv \sim p_{\theta_1}(\cdot)$ and then generate $\xv \sim p_{\theta_2}(\cdot \mid \yv, \zv)$.

In practice, we observe $\zv$ and $\yv$ become more and more decorrelated during the training, but this happens at a slow pace.
We can further enforce this decorrelation using a mini-batch cross-correlation penalty  
$$
C(\{\yvh\}, \{\zv\}) \defeq \frac{1}{2} \sum_{i,j} \big ( \frac{1}{N} \sum_{n=1}^{N}(\yvh_i\ttt{n} - \bar{\yv}_i)(\zv_i\ttt{n} - \bar{\zv}_i) \big)^2
$$
where $\{\yvh\}$/$\{ \zv\}$ are conditions/codes in a \textit{mini-batch} of size $N$, where $\zv \sim q_\psi(\cdot \mid \xvh)$ and $i,j$ index dimensions of $\yvh, \zv$ respectively. Here $\bar{\yv}_i$ and $\bar{\zv}_i$ are mini-batch average values. 

\begin{figure}[ht]
  \centering
\includegraphics[width=1\linewidth]{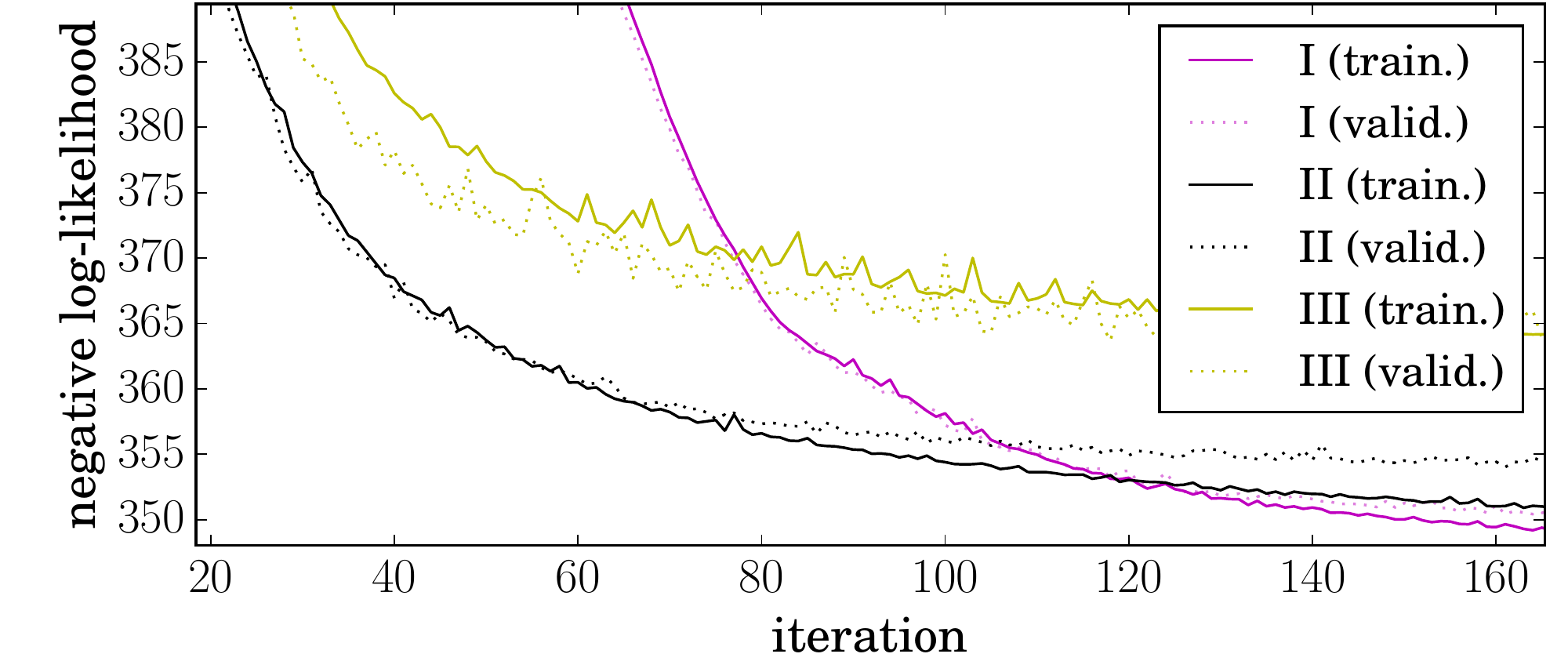} 
  \caption{{\small Negative log-likelihood of different C-VAE schemes. Note that scheme II can only serve as a baseline and due to correlation between $\yvh$ and $z$ cannot be used for conditional sampling.}}
\label{fig:vae_nll}
\end{figure}

Lack of cross-correlation only entails independence, if both $\yv_i$ and $\zv_i$ have Gaussian distribution. Although $p_{\theta_1}(\zv)$ is by design a standard Gaussian, the condition $y$ may have 
an arbitrary distribution. To resolve this, we transform $\yvh_i \to {F}^{-1}_{\mathcal{N}}({F}_{y_i}(\yvh_i))$, where ${F}_{y_i}$ is the empirical cumulative distribution function (CDF) for $\yvh \in \DD$ and $F^{-1}_{\mathcal{N}}$ is the (numerically approximated) inverse CDF of Gaussian. The transformed variable has a Gaussian distribution.

\subsection{Experiments}
\Cref{fig:cc} compares the reduction in the average cross-correlation between ${\yvh}$ and ${\zv}$ for the same network, with and without the cross-correlation penalty. For numerical stability we linearly increase the penalty coefficient from 0 to 1000 over iterations.
These results are for the \COSMOS dataset. All C-VAE results are using the log-pixel-intensity, also for numerical stability.

\Cref{fig:vae_nll} compares $-\log(p_{\theta}(\xvh \mid \yvh))$ for three models: 
\begin{enumerate}[I]
\item using a neural network to encode $p_{\theta_1}(\zv \mid \yv)$ 
\item using $p_{\theta_1}(\zv \mid \yv) = p_{\theta_1}(\zv)$ 
\item $p_{\theta_1}(\zv \mid \yv) = p_{\theta_1}(\zv)$ plus cross-correlation penalty
\end{enumerate}
The figure suggests that the first scheme eventually produces better models. It also shows that 
enforcing the independence of $\zv$ and $\yv$ only slightly decreases the likelihood, compared to the baseline II where $\zv$ and $\yv$ remain highly dependent.

\section{A New Objective for Adversarial Training}\label{sec:cgan}
A major problem with VAE-generated images is their blurriness. A few recent works address this issue~\cite{KingmaSW16,larsen2015autoencoding,dosovitskiy2016generating} -- \eg by defining a more expressive reconstruction loss. 
Fortunately, the noise model is available for \COSMOS images,
and the added noise to some extent reduces this problem in our application (see \cref{sec:validation}).

\begin{figure}[ht]
  \centering
\includegraphics[width=1\linewidth]{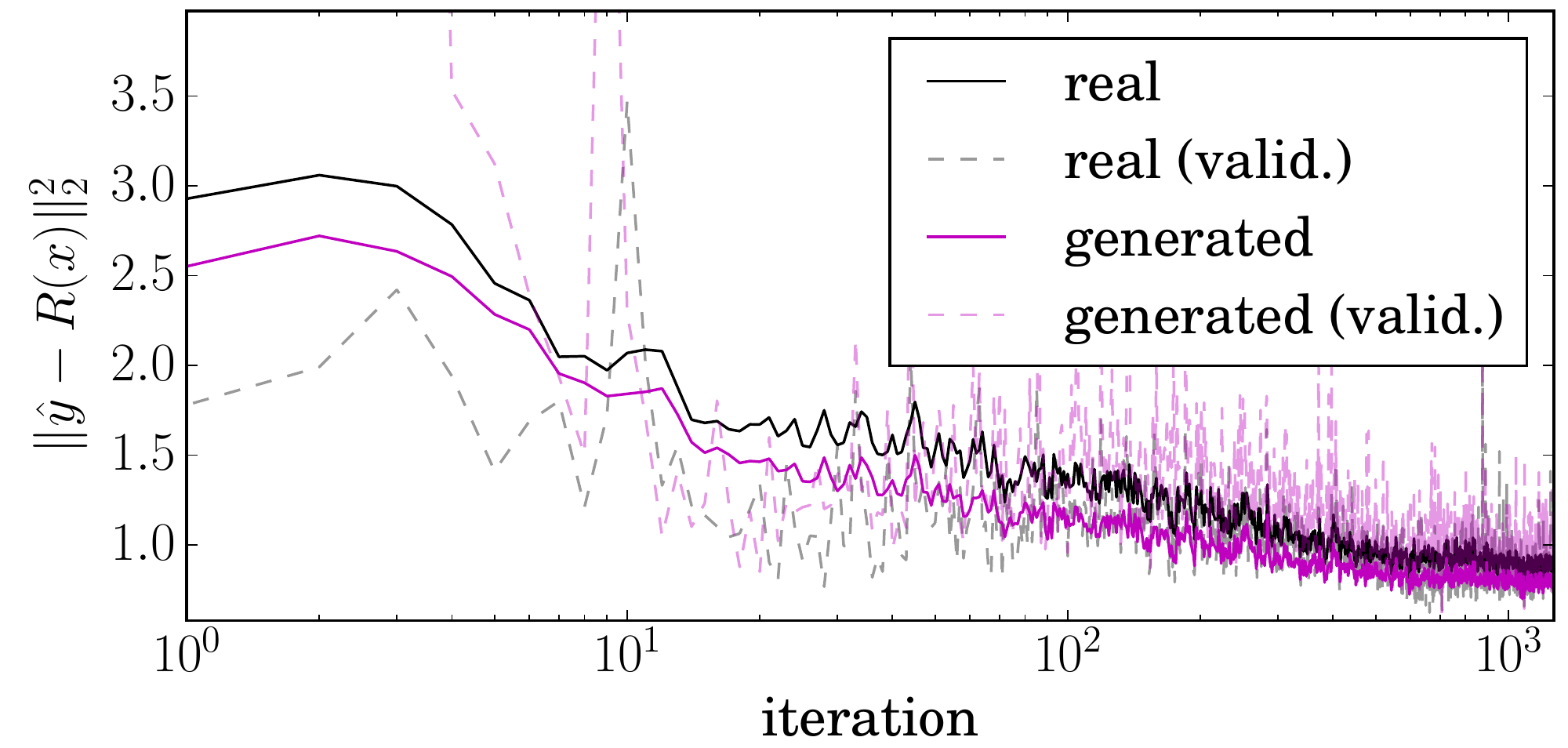} 
\caption{{\small 
The prediction error for real and generated images in C-GAN for \COSMOS dataset.
 }}
\label{fig:adv_rf}
\end{figure}

An alternative to generative modeling that does not suffer from this problem is offered by adversarial training of generative networks~\cite{goodfellow2014generative}.
In the \textbf{adversarial setting}, a generator $G_\omega:\ZZ \to \XX$ attempts to fool the discriminator $D_{\psi}:\XX \to [0,1]$ into classifying its fake instances $\xv = G(\zv)$ as real, while the discriminator's objective is to correctly classify the two sources of real versus generated instances.
Deep networks representing these adversaries are trained alternatively, and under some conditions $p_G$ (the implicit distribution of the generator $G_\omega$ for $\zv \sim \mathcal{U}(0,1)$) converges to $p^*$ --\ie at this fixed-point, the generator produces realistic images that are indistinguishable by the discriminator.

The \textbf{conditional variation} of this method was first introduced by~\citet{mirza2014conditional} and used in a cascade of conditional models with increasing resolution in \cite{denton2015deep}. 
In these conditional models, the generator $G_\omega:\ZZ \times \YY \to \XX$ and the discriminator $D_\psi:\XX \times \YY \to [0,1]$,
are both deep neural networks that are now conditioned on the same observed variable $\yvh \in \DD$. The min-max formulation of this adversarial setting seeks a saddle-point for
\begin{align*}
  \min_{\omega}\, \max_{\psi} \quad &\Ex_{\xvh, \yvh \in \DD, \zv \in \mathcal{U}}\, \big [\log(D_{\psi}(\xvh, \yvh))\\ 
+ &\log(1 - D_{\psi}(G_{\omega}(\zv, \yvh), \yvh)) \big ]
\end{align*}

In practice it is much more efficient to use a different loss function for the generator as it produces stronger gradients for the generator at the beginning~\cite{goodfellow2014generative}:
\begin{align*}
\max_{\psi} \quad &\Ex_{\xvh, \yvh \in \DD, \zv \in \mathcal{U}}\, \big [ \log(D_{\psi}(\xvh, \yvh)) + \\ 
& \log(1 - D_{\psi}(G_{\omega}(\zv, \yvh), \yvh)) \big ] \\
\max_{\omega} \quad &\Ex_{\zv \in \mathcal{U}}\, \big[ \log(D_{\psi}(G_{\omega}(\zv, \yvh), \yvh)) \big ]
\end{align*}

Here, one must carefully adjust the expressive power of $G$ and $D$ to avoid oscillations, and domination of either adversary. The choice of hyper-parameters is known to be a major hurdle in training of adversarial networks and using this scheme, despite much effort, we could not train a generator for our problem that uses \textit{continuous} conditional variables.

\begin{figure}
\includegraphics[width=\columnwidth]{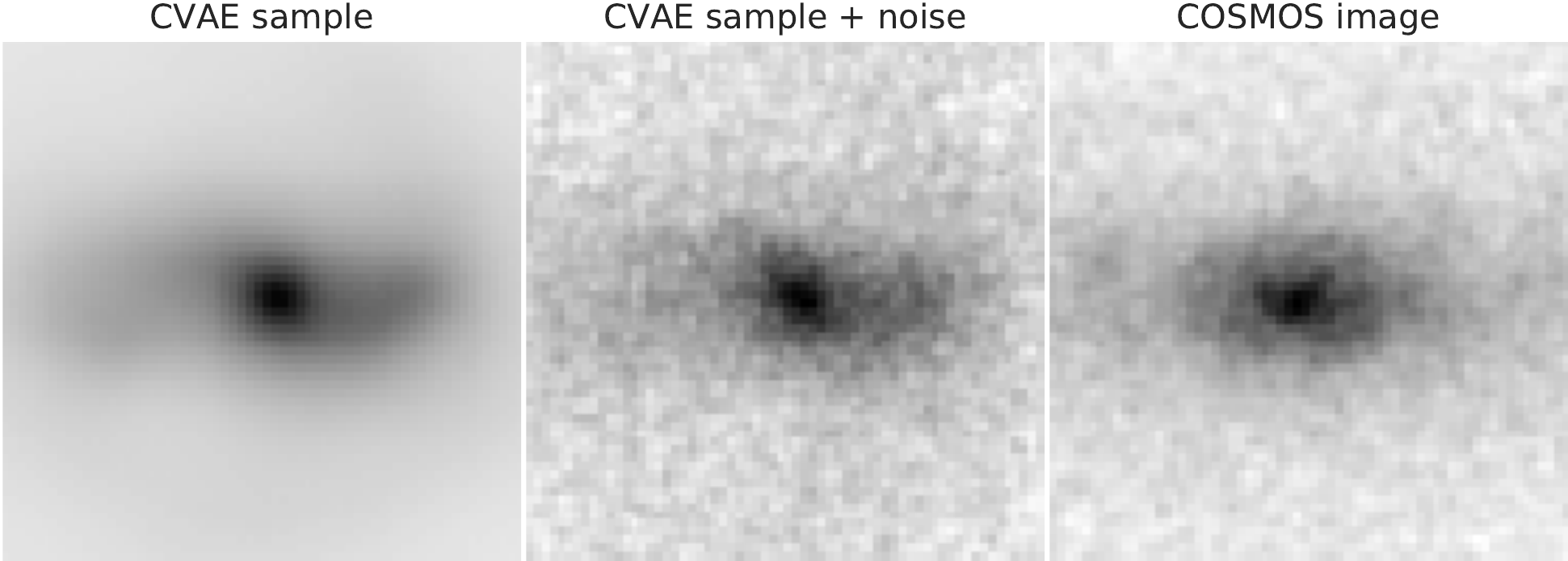}
\caption{{\small Comparison of a C-VAE sample before and after adding noise and a real \COSMOS image with corresponding size, magnitude and redshift.}}\label{fig:noise}
\end{figure}

We introduce an \textbf{alternative adversarial objective} for conditional
generative modeling that in our experience is more stable and did not require any hyper-parameter tuning in our application.
The basic idea is simple: A \textbf{predictor} $R: \XX \to \YY$ replaces the discriminator $D: \XX \times \YY \to [0,1]$. The predictor attempts to produce predictions of the condition $\yvh \in \DD$ for the real data, that are at least as good as its predictions for generated instances. The generator's objective is to produce instances with low prediction error


\begin{align}
&\text{Predictor:} \quad \min_{\psi}  \quad \min \{0, \label{eq:pred_cont}\\
&\Ex_{\xvh, \yvh \in \DD, \zv \in \mathcal{U}} \, \big [ \ell(R_{\psi} (G_{\omega}(\zv, \yvh)), \yvh) - \ell(R_{\psi}(\xv), \yvh) \big ] \} \notag \\
&\text{Generator:} \quad  \min_{\omega} \quad \Ex_{\yvh \in \DD, \zv \in \mathcal{U}}\,\big [ \ell(R_{\psi}(G_{\omega}(\zv, \yvh)), \yvh) \big] \label{eq:gen_cont}
\end{align}
where in our application $\ell(\yv, \yvh) = \|\yv - \yvh \|_2^2$. 

\begin{figure}[ht]
    \centering
    \hbox{
    \begin{subfigure}[b]{0.4\linewidth}
        \includegraphics[width=\textwidth]{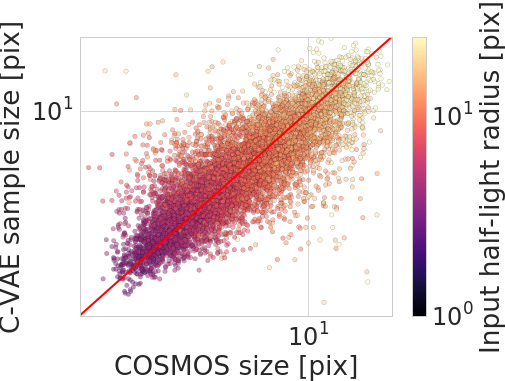}
        \caption{Galaxy sizes}
        \label{fig:comp_size}
    \end{subfigure}
    ~\qquad
    \begin{subfigure}[b]{0.4\linewidth}
        \includegraphics[width=\textwidth]{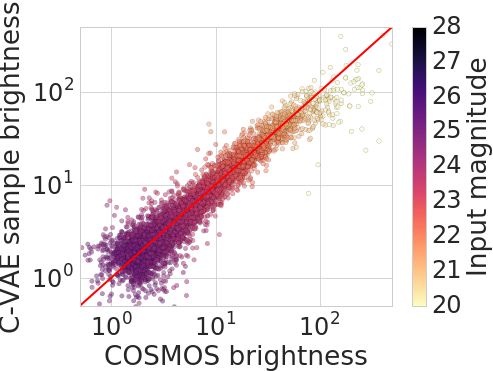}
        \caption{Galaxy brightness}
        \label{fig:comp_bright}
    \end{subfigure}
    }
    \caption{{\small Comparison of galaxy sizes and brightness between real \COSMOS images and C-VAE samples. Colors indicate the value of the relevant variable used to condition the generated images (half-light radius for size and magnitude for brightness)}}\label{fig:comparison}
\end{figure}

Why should the generator produces realistic images at all as long as the predictor makes equally bad predictions for both real and generated images? Both errors \cref{eq:pred_cont,eq:gen_cont} will be low in this case. The key here is that the generator always seeks to improve its samples to increase their prediction accuracy and therefore the dynamics of this adversarial setting does not allow this mode of failure.

This scheme, also relaxes the constraint on the \textbf{expressive power of the adversaries}. This is because
the predictor has no incentive to lower the error for the real data, as long as its prediction errors are not worse that those of the generated data. Therefore, it is only the generator that fuels the competition and training is practically finished when the generator is unable to improve.

A mode of failure that our scheme does not resolve is the \textbf{collapse of generator}, where generator $G(\yv, \zv)$ repeats few output patterns by solely relying on $\yv$ and basically ignoring the random feed $\zv$. The predictor eventually realizes this repeating pattern in generated data but gradient descent can no longer rescue the generator from this local optima. A solution to this problem called mini-batch discrimination was recently proposed by \citet{salimans2016improved}, where each instance in the mini-batch is augmented with information about its differences with other instances in the same mini-batch. The Predictor can therefore detect this tendency of the generator early on, and the generator incurs a loss for its behavior before its complete collapse. For better mini-batch statistics, we use relatively larger mini-batches with 128/256 instances.

\subsection{Experiments}
Following~\citet{radford2015unsupervised} we use (de)convolutional layers with (fractional) stride, batch normalization~\cite{ioffe2015batch} and leaky-ReLU activation functions in our deep networks.
For optimization, we use Adam~\cite{kingma2014adam} with reduced exponential decay rate of .5 for the first moment estimates.

\Cref{fig:adv_rf} reports the prediction loss $\ell(R_{\psi} (G_{\omega}(\zv, \yvh), \yvh))$ and $\ell(R_{\psi}(\xv), \yvh)$ for the \COSMOS dataset, were we use 4 (de)convolution layers.
The figure suggests that the predictor tends to keep the prediction error of the 
real images slightly higher than that of generated images. Both of these quantities reduce over time,
and their agreement with validation errors could monitor convergence. The fact that the error is decreasing over time and prediction error for both real and generated data remains close to each other
is due to having a ``laid back'' predictor -- \ie by removing the $\min(0,.)$ operation in predictor's loss, we would lose both of these properties. 

For illustration purposes, we applied the same method to the \GALAXYZOO dataset. \Cref{fig:galaxyzoo} shows some instances in the test-set accompanied by C-GAN generated image conditioned on the same $\yvh$. For this dataset we used 5-layer fully (de)convolutional generator and predictor, mini-batch discrimination, batch-normalization and $\tanh$ activation function for the final layer of the generator.

\section{Validation}\label{sec:val}\label{sec:validation}
In this section, we assess the quality of the model generated galaxy images by comparing common image statistics used in weak lensing analyses. Our aim is to consistently measure the same statistics on real \COSMOS images and images generated by our model for the same set of input variables $y$. These statistics are affected by the presence of noise in the image, but as was noted in the previous section, our C-VAE generates essentially noiseless images, which prevents direct comparison with real images. We  limit this analysis to C-VAE generated images (as we found it to produce more consistent results compared to C-GAN) and add a noise field to our generated images. This noise model, calibrated for \COSMOS observations, is provided by the GalSim package; see \cref{fig:noise}.

The most commonly used image statistics in weak lensing analyses rely on the second moments of the  galaxy's intensity profile $I(u_1, u_2)$, where $ ( u_1, u_2)$ are pixel coordinates. The second moment tensor ${Q}$ 
is defined as:
\begin{equation*}
	Q_{\alpha \beta}= \frac{\int \mathrm{d}u_1 \mathrm{d}u_2 \ W(u_1, u_2) \  I(u_1, u_2) \ u_\alpha u_\beta}{ \int \mathrm{d}u_1 \mathrm{d}u_2 \ W(u_1, u_2)  \ I(u_1 , u_2) },
\end{equation*}
with $(\alpha,\beta ) \in \{ 1, 2\}$ and where $W$ is a weighting function. This tensor can be used to define a size measurement $\sigma = |\det( {Q} )|^{1/4}$ which reduces to the standard deviation if the light profile is a Gaussian. More importantly, the second moments are commonly used to measure galaxy ellipticities which can be defined as:
\begin{equation*}
	e = e_1 + i e_2 = \frac{Q_{11} - Q_{22} -2i Q_{12}}{Q_{11} + Q_{22} + 2(Q_{11} Q_{22} - Q_{1 2}^2)^{1/2}}
\end{equation*}
To measure ${Q}$ in practice, we use the \textit{adaptive moments} method \cite{Hirata2003,Mandelbaum2005} which estimates the second order moments by fitting an elliptical Gaussian profile to the galaxy light profile. As a side product of this method, we can also use the amplitude of the best fit Gaussian model as a proxy for the brightness of the  galaxy.

\begin{figure}[ht]
  \centering
    \begin{subfigure}[b]{0.49\columnwidth}
        \includegraphics[width=\textwidth]{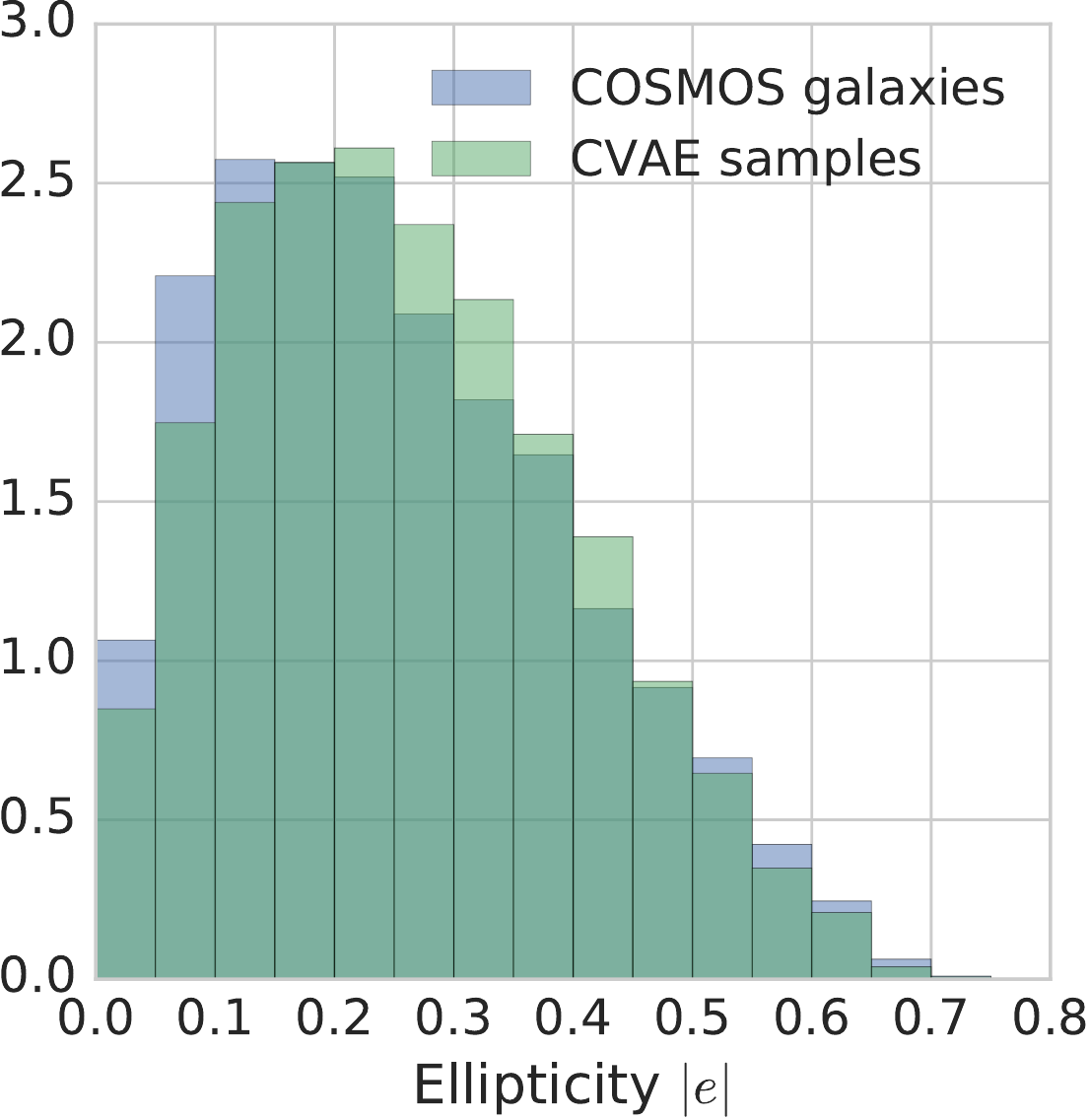}
        \label{fig:ellipticity_distribution}
    \end{subfigure}~
    \begin{subfigure}[b]{0.49\columnwidth}
        \includegraphics[width=\textwidth]{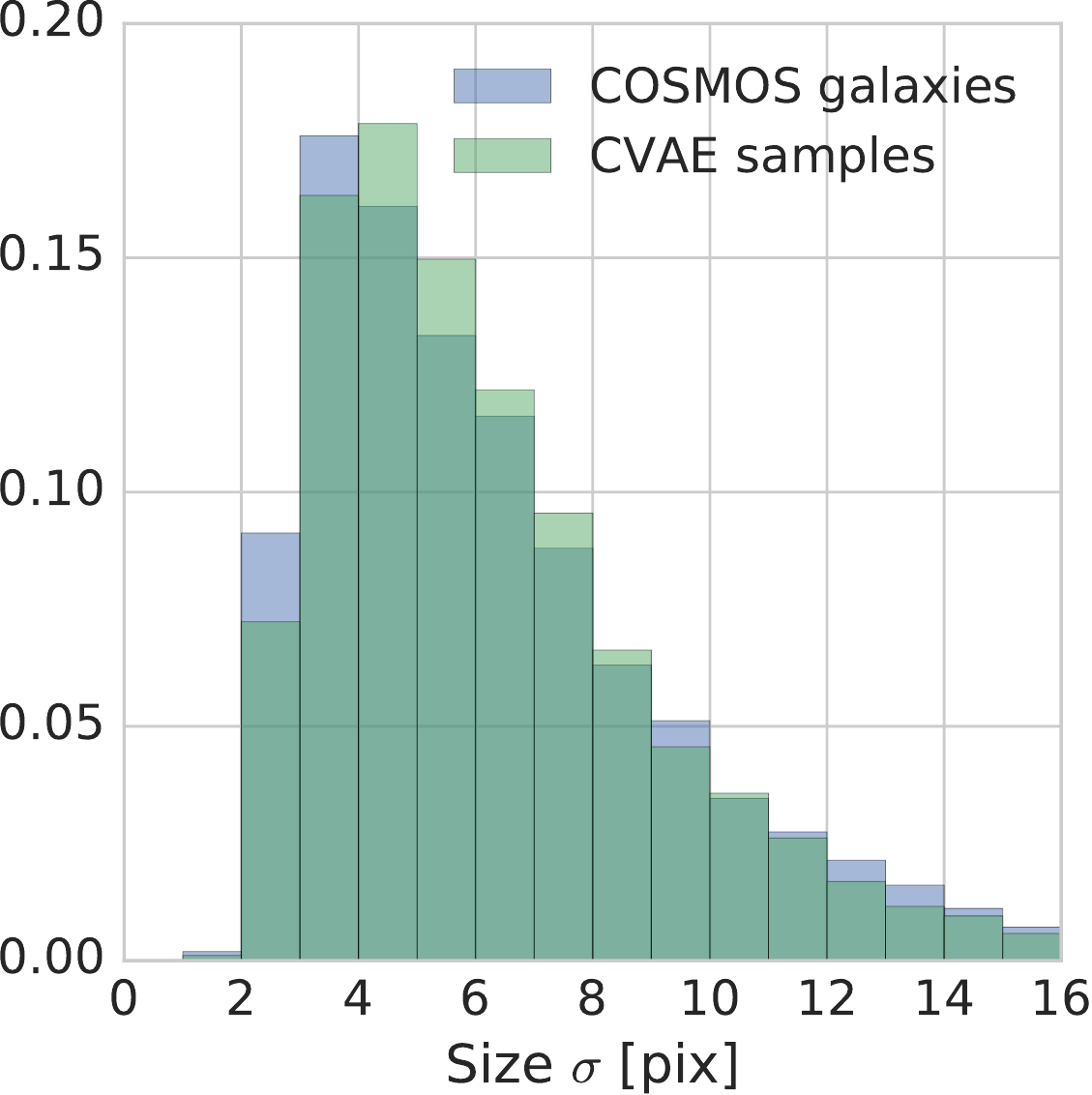}
        \label{fig:size_distribution}
    \end{subfigure}
    \caption{{\small Comparison of galaxy ellipticity (left) and size (right) distributions measured from second moments between real \COSMOS images and CVAE samples.}}
\label{fig:distributions}
\end{figure}

We compare real \COSMOS images to C-VAE samples by processing the images in pairs, where every \COSMOS galaxy in our validation set is associated to a C-VAE sample conditioned on the half-light radius, magnitude and redshift of the real galaxy.
\cref{fig:comp_size} shows for each pair of images the galaxy size $\sigma$, as measured using second order moments; see also \cref{fig:cosmos_images}. The color of the points indicates the half-light radius of the \COSMOS galaxy in the pair, also used to condition the C-VAE sample. 
As can be seen, the sizes of generated galaxies are generally unbiased. 
\cref{fig:comp_bright} shows the similar results for brightness; C-VAE is generating samples of the correct brightness without any significant bias.

The most relevant image statistics for weak lensing science are the ellipticity and size distributions of a given galaxy sample. \cref{fig:distributions} compares these overall distributions measured on real and generated galaxies. Note that contrary to the previous test where the quantities considered (size and brightness) were part of the condition variable $y$, the ellipticity is not. Therefore, this test allows us to check how well the model is able to blindly learn correct galaxy shapes. This figure shows that despite being slightly more elliptical than real galaxies, the ellipticity distribution of the C-VAE samples is broadly consistent with the \COSMOS distribution. \cref{fig:distributions} also compares size distributions which are in good agreement. This comes as no surprise however as C-VAE samples are explicitly conditioned on galaxy sizes and the previous test has shown these samples to be largely unbiased.

\section*{Conclusion}
In this paper, we proposed novel techniques and studied the application of two most promising methods for deep conditional generative modeling in producing galaxy images. In the future, we plan to measure more subtle morphological statistics in generated images and find ways for simultaneous learning of the noise model. We are also investigating the application of our variation on adversarial training in other settings and assessing the effectiveness of the predictor as a stand-alone classification/regression model.

\bibliographystyle{plainnat}
\bibliography{refs}

\newpage
\appendix
\section*{C-GAN results}

\begin{figure}[ht]
    \centering
    \hbox{
    \begin{subfigure}[b]{0.4\linewidth}
        \includegraphics[width=\textwidth]{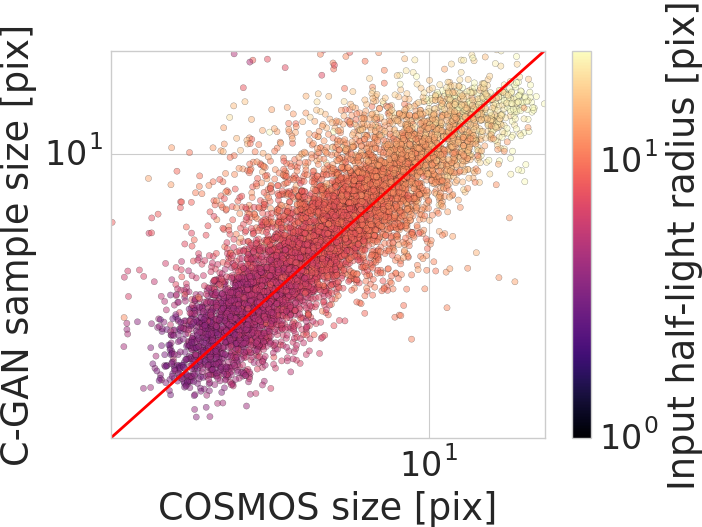}
        \caption{Galaxy sizes}
        \label{fig:cgan_comp_size}
    \end{subfigure}
    ~\qquad
    \begin{subfigure}[b]{0.4\linewidth}
        \includegraphics[width=\textwidth]{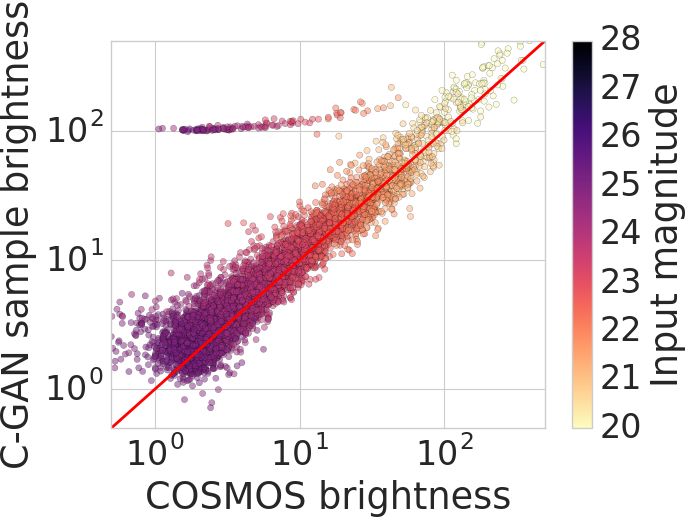}
        \caption{Galaxy brightness}
        \label{fig:cgan_comp_bright}
    \end{subfigure}
    }
    \caption{{\small Comparison of galaxy sizes and brightness between real \COSMOS images and C-GAN samples. Colors indicate the value of the relevant variable used to condition the generated images (half-light radius for size and magnitude for brightness)}}\label{fig:cgan_comparison}
\end{figure}

This appendix complements Section~\ref{sec:val} by providing the results of the C-GAN on the \COSMOS data. Following the same approach as for the validation of the C-VAE results, we first compare size and brightness statistics measured on pairs of real \COSMOS images and C-GAN samples, with same conditional values. As can be seen in Fig.~\ref{fig:cgan_comparison}, the size and brightness of the galaxies generated by the C-GAN are largely unbiased and similar to the C-VAE results. 

\begin{figure}[ht]
  \centering
    \begin{subfigure}[b]{0.49\columnwidth}
        \includegraphics[width=\textwidth]{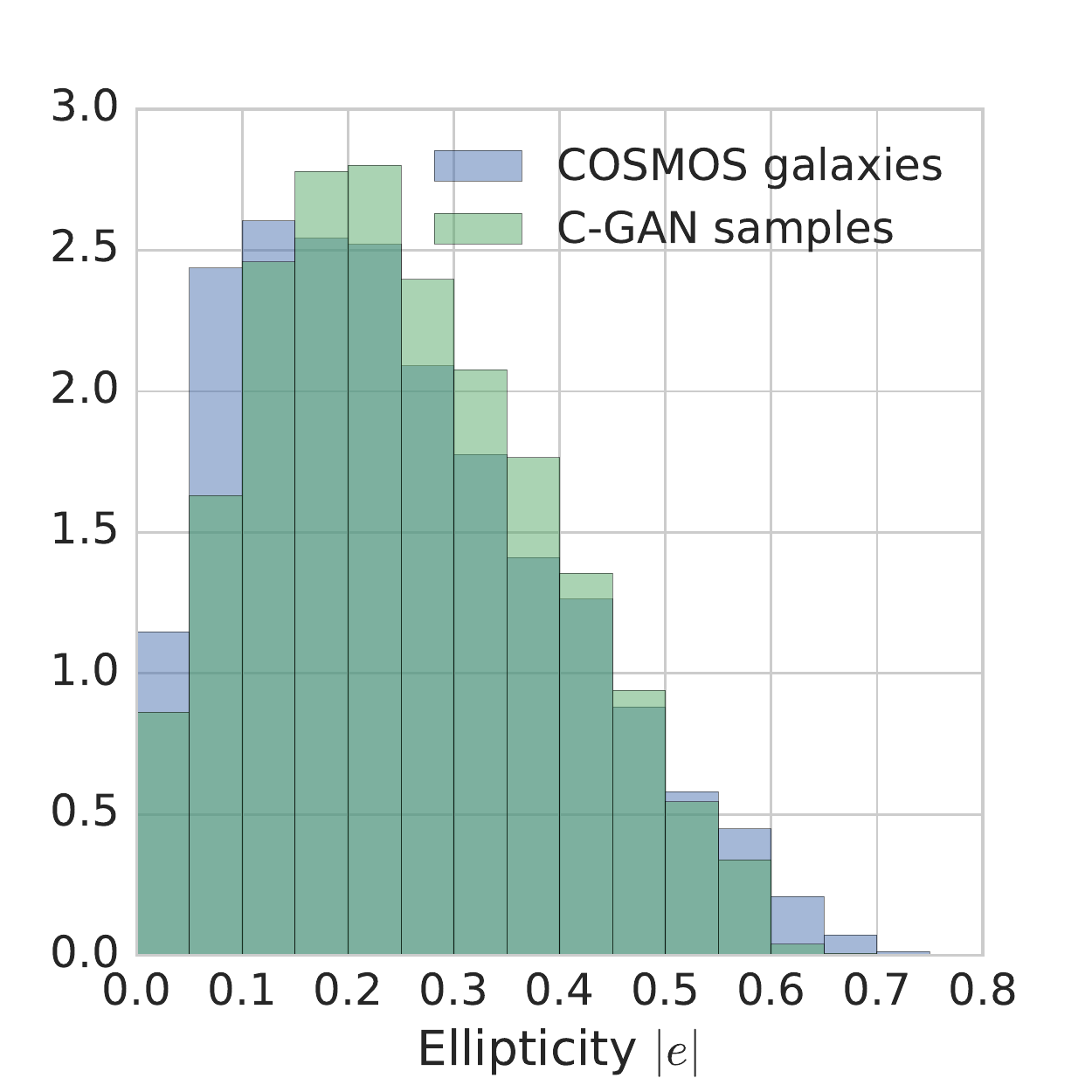}
        \label{fig:cgan_ellipticity_distribution}
    \end{subfigure}~
    \begin{subfigure}[b]{0.49\columnwidth}
        \includegraphics[width=\textwidth]{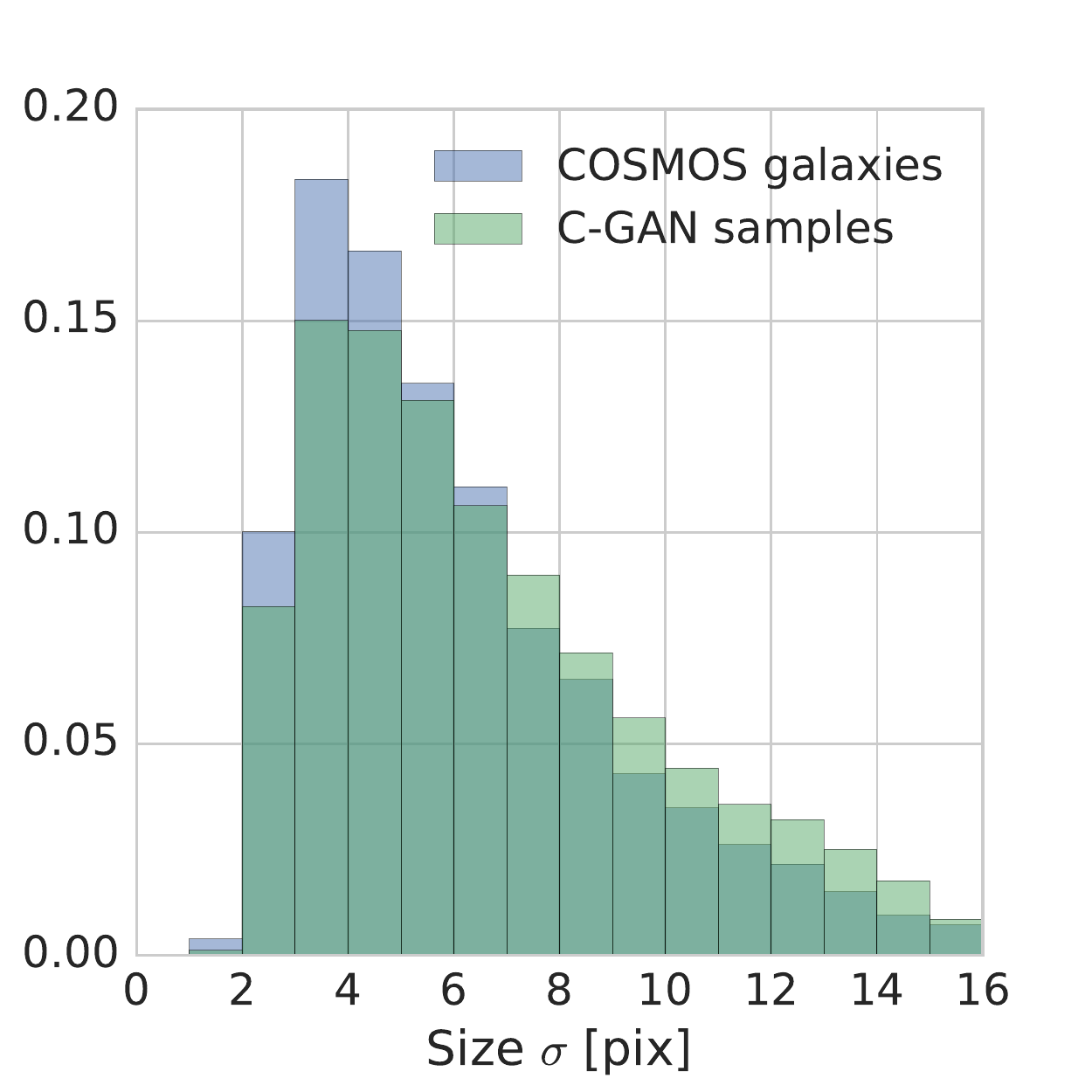}
        \label{fig:cgan_size_distribution}
    \end{subfigure}
    \caption{{\small Comparison of galaxy ellipticity (left) and size (right) distributions measured from second moments between real \COSMOS images and C-GAN samples.}}
\label{fig:cgan_distributions}
\end{figure}

\end{document}